\documentclass[superscriptaddress,aps,prx,twocolumn,showpacs,nofootinbib, longbibliography, notitlepage,floatfix]{revtex4-2}
\usepackage{etex}
\usepackage{amsmath,amssymb,amsthm}
\usepackage[colorlinks=true,citecolor=blue,urlcolor=blue]{hyperref}
\usepackage[pdftex]{graphicx}
\usepackage{times,txfonts}
\usepackage{braket}
\usepackage{color}
\usepackage{natbib}
\usepackage{amsmath,blkarray}
\usepackage{mathtools}
\usepackage{ulem}
\usepackage{latexsym}
\usepackage{tabularx, booktabs}
\usepackage{graphics,epstopdf}
\usepackage{graphicx}
\usepackage{float}
\usepackage{amsfonts}
\usepackage{tikz}
\usetikzlibrary{quantikz}
\usepackage{color,soul}

\newcommand{\be}{\begin{equation}}
\newcommand{\ee}{\end{equation}}
\newcommand{\ba}{\begin{eqnarray}}
\newcommand{\ea}{\end{eqnarray}}

\usepackage{multirow}
\usepackage{appendix}
\usepackage{url}

\begin{document}
\title{Photonic counterdiabatic quantum optimization algorithm }

\author{Pranav Chandarana}
\email{pranav.chandarana@gmail.com}
\affiliation{Department of Physical Chemistry, University of the Basque Country UPV/EHU, Apartado 644, 48080 Bilbao, Spain}
\affiliation{EHU Quantum Center, University of the Basque Country UPV/EHU, Barrio Sarriena, s/n, 48940 Leioa, Spain}

\author{Koushik Paul}
\affiliation{Department of Physical Chemistry, University of the Basque Country UPV/EHU, Apartado 644, 48080 Bilbao, Spain}
\affiliation{EHU Quantum Center, University of the Basque Country UPV/EHU, Barrio Sarriena, s/n, 48940 Leioa, Spain}

\author{Mikel Garcia-de-Andoin}
\affiliation{Department of Physical Chemistry, University of the Basque Country UPV/EHU, Apartado 644, 48080 Bilbao, Spain}
\affiliation{EHU Quantum Center, University of the Basque Country UPV/EHU, Barrio Sarriena, s/n, 48940 Leioa, Spain}
\affiliation{TECNALIA, Basque Research and Technology Alliance (BRTA), 48160 Derio, Spain}

\author{Yue Ban}
\affiliation{TECNALIA, Basque Research and Technology Alliance (BRTA), 48160 Derio, Spain}

\author{Mikel Sanz}
\affiliation{Department of Physical Chemistry, University of the Basque Country UPV/EHU, Apartado 644, 48080 Bilbao, Spain}
\affiliation{EHU Quantum Center, University of the Basque Country UPV/EHU, Barrio Sarriena, s/n, 48940 Leioa, Spain}
\affiliation{IKERBASQUE, Basque Foundation for Science, 48009 Bilbao, Spain}
\affiliation{Basque Center for Applied Mathematics BCAM, 48009 Bilbao, Spain}

\author{Xi Chen}
\email{chenxi1979cn@gmail.com}
\affiliation{Department of Physical Chemistry, University of the Basque Country UPV/EHU, Apartado 644, 48080 Bilbao, Spain}
\affiliation{EHU Quantum Center, University of the Basque Country UPV/EHU, Barrio Sarriena, s/n, 48940 Leioa, Spain}

\begin{abstract}
We propose a hybrid quantum-classical approximate optimization algorithm for photonic quantum computing, specifically tailored for addressing continuous-variable optimization problems. Inspired by counterdiabatic protocols, our algorithm significantly reduces the required quantum operations for optimization as compared to adiabatic protocols. This reduction enables us to tackle non-convex continuous optimization and countably infinite integer programming within the near-term era of quantum computing. Through comprehensive benchmarking, we demonstrate that our approach outperforms existing state-of-the-art hybrid adiabatic quantum algorithms in terms of convergence and implementability. Remarkably, our algorithm offers a practical and accessible experimental realization, bypassing the need for high-order operations and overcoming experimental constraints. We conduct proof-of-principle experiments on an eight-mode nanophotonic quantum chip, successfully showcasing the feasibility and potential impact of the algorithm.
\end{abstract}

\maketitle
\section{Introduction}
Harnessing usefulness from current noisy intermediate-scale quantum (NISQ)~\cite{Preskill2018quantumcomputingin} computers has emerged as the main objective of the quantum computing community. Variational quantum algorithms (VQAs) are the leading candidates to achieve this goal, making use of the limited quantum resources that existing NISQ computers offer~\cite{Cerezo2021}. These hybrid algorithms aim to solve computationally demanding tasks with enhanced efficiency by synergistically combining the computational power of classical systems. In the past few years, VQAs have demonstrated significant potential in addressing numerous challenges in contemporary science, such as problems involving many-body quantum Hamiltonians~\cite{PhysRevB.102.235122,BravoPrieto2020scalingof}, quantum chemistry~\cite{PhysRevX.8.011021,Adaptvqe,Kandala2017}, combinatorial optimization~\cite{10.1007/978-3-030-14082-3_7,Karamlou2021}, and others~\cite{Robert2021,10.1145/3520304.3533986}.

The fundamental structure of any VQA typically consists of three essential elements; encoding, processing, and decoding. The first step involves transforming a logical optimization problem into a Hamiltonian, encoding the solution in its ground state. 
The processing phase refers to designing a parameterized quantum circuit or the circuit ansatz to manipulate quantum states, whereas the decoding step includes a measurement scheme to evaluate the cost function. This cost function is minimized using classical optimization routines that eventually produce an optimal quantum state.  
Currently, most research efforts are focused on encoding discrete optimization problems using qubit-based approaches, which are well-suited for implementation in superconducting circuits and trapped ions. 

In contrast, photonic quantum computing (PQC) 
incorporates the continuous variable (CV) formalism. This allows quantum information to be encoded in the quadrature amplitudes of an electromagnetic field~\cite{PhysRevResearch.1.033063} known as qumodes. This encoding provides benefits in representing continuous optimization problems that are expensive to encode with qubits. 
Recently, there has been a growing interest in developing CV quantum algorithms for solving various problems in the PQC paradigm. For instance, a CV-based quantum approximate optimization algorithm (CV-QAOA) has recently been proposed and benchmarked with the minimization of the non-convex Styblinski-Tang function~\cite{verdon2019quantum}. Likewise, a CV adiabatic quantum algorithm was also proposed that investigated mixed-integer programming problems using Fock  encoding~\cite{khosravi2021mixed}. In addition, efforts have been made to encode graph problems, imaginary time evolution for quantum field theories, Grover search and instantaneous quantum polytime circuits over continuous spaces~\cite{mezher2023solving, PhysRevA.105.012412, pati2000quantum,PhysRevLett.118.070503}. Apart from optimization, the CV regime is also utilized as a tool for error-correcting codes~\cite{PhysRevA.64.012310,Terhal_2020} and for quantum state learning~\cite{Arrazola_2019}.


One of the challenges encountered while implementing adiabatic algorithms using the CV approach is the inherent bottleneck arising from the quadratic nature of Gaussian operations in phase space. Generally, complex optimization problems require non-Gaussian operations and intricate gate decomposition for effective time evolution. 
The experimental feasibility of these algorithms in solving high-degree problems is limited, as recent experiments have only been able to solve simple quadratic function of $x$ in the case of CV-QAOA~\cite{enomoto2022continuous}. Therefore, there is a pressing need for algorithms that operate within low degrees while still enabling experimental exploration of complex high-degree problems using near-term photonic devices. 

In this article, we propose the photonic counterdiabatic quantum optimization (PCQO) algorithm to address this challenge. PCQO is a hybrid quantum-classical algorithm designed to solve problems suitable for currently available photonic devices by utilizing a circuit ansatz and a classical optimization routine. This circuit ansatz is designed from a pool of Gaussian and non-Gaussian operations that are obtained by drawing inspiration from counterdiabatic (CD) protocols~\cite{PhysRevLett.111.100502}. These CD protocols, known as shortcuts-to-adiabaticity methods~\cite{insta9,sta16}, accelerate the adiabatic process and circumvent the typically slow evolution mandated by the adiabatic theorem~\cite{PhysRevA.83.062116}. Previous applications of these methods have demonstrated substantial improvements in QAOA and digitized adiabatic evolution~\cite{PhysRevResearch.4.013141,nar,PhysRevA.104.L050403,PhysRevResearch.4.043204,PhysRevResearch.4.L042030}.

We investigate the performance of this algorithm for (a) phase-space encoding, which encodes information in the $\hat{x}$ quadrature, and (b) Hilbert-space encoding, which represents information using Fock states $\ket{n}$. The former includes classical non-convex continuous optimization problems~\cite{Rosenbrock1960,STYBLINSKI1990467} and the latter includes integer programming problems~\cite{KOLMAN1995249}. Additionally, we show that PCQO outperforms state-of-the-art quantum algorithms such as CV-QAOA in terms of performance and implementability. Lastly, we provide considerations for implementing PCQO with NISQ devices and show a proof-of-principle experiment that solves a simple problem in an eight-mode photonic quantum computer.

This article is structured as follows. Section.~\ref{sec:var_quantum} provides a discussion of variational quantum algorithms in the context of qubits and qumodes. In Section.~\ref{sec:counterdiabatic}, we present a detailed description of the PCQO algorithm and benchmarked the algorithm in Section.~\ref{sec:phase_space} for non-convex continuous optimization problems using phase-space encoding. In Section.~\ref{sec:hilbert_space}, we apply PCQO to countably infinite discrete optimization like unbounded knapsack problem and Maxclique problem using Hilbert-space encoding. Section. \ref{sec:experimental} provides a proof-of-principle experiment to further establish the feasibility of our algorithm in the near-term era. Finally, we conclude in Section.~\ref{sec:discussion} and discuss possible future directions.

\section{Variational quantum algorithms}\label{sec:var_quantum}
In the qubit-based regime, the encoding is generally done by considering a physical system such as a molecule or a spin chain. A problem Hamiltonian $H_p$ corresponding to this system is found such that its ground state entails the information of the solution. The processing phase employs a circuit ansatz comprising various parameterized gates. The design of this ansatz is critical to the performance of the VQA, as it directly affects the energy landscape and consequently, the convergence and success rate of the optimization process. 

Circuit ansatzes can be broadly divided into two categories: problem-inspired and hardware efficient. In problem-inspired ansatzes, the parameterized unitary is constructed by taking information from the $H_p$ which usually corresponds to time evolutions. For example, in the quantum approximate optimization algorithm (QAOA), the parameterized unitary is given by $U(\alpha)= e^{-i\alpha H_p}$, where $\alpha$ is an optimizable parameter~\cite{farhi2014quantum}. On the other hand, the hardware-efficient ansatzes are specifically designed to take into account the connectivity constraints of the underlying quantum hardware~\cite{Kandala2017}. Both problem-inspired and hardware-efficient ansatzes have distinct roles in VQA design, providing varying trade-offs between performance and feasibility in practical implementations. The choice of ansatz depends on factors like the problem characteristics, available resources, and desired accuracy. 

Ideally, problem-inspired ansatzes should be prioritized over hardware-efficient ansatzes because they narrow down the search space, enhancing trainability. However, problem-inspired ansatzes often result in increased circuit depths and unfavorable connections. Consequently, there is significant interest in developing algorithms containing problem-inspired ansatzes that can be implemented on near-term devices~\cite{blekos2023review}.

Lastly, the decoding step involves measurements, typically in the computational basis, to evaluate a cost function. This cost function maps the optimizable parameters to real numbers. In many cases, the cost function corresponds to the expectation value of the $H_p$. However, alternative metrics such as fidelity or conditional value at risk~\cite{Barkoutsos2020improving} can also be considered.

In the qumode-based regime, the components required for designing a VQA differ intrinsically from those in the qubit-based regime. In the encoding step, we have the flexibility to choose between two formulations: the phase space formulation and the Hilbert space formulation. These correspond to the wave-like and particle-like nature of light, respectively. In the phase space picture, the state of a single qumode is represented using $(\hat{x},\hat{p})$, which are the position and momentum operators, respectively. The problem Hamiltonian can be expressed as $H_p=F(\hat{x},\hat{p})$. On the other hand, qumode states can also be represented in a countably infinite Hilbert space spanned by Fock states $\ket{n_{i=0,1,\dots}}$. Consequently, the problem Hamiltonian can be written as $H_p = F(\hat{n})$. 

In the processing stage, the overall recipe remains the same, but PQC involves different operations designed for qumodes. These operations can be categorized into Gaussian operations and non-Gaussian operations. As the name suggests, Gaussian operations map Gaussian states to themselves and are generated by quadratic operations in $\hat{x}$ and $\hat{p}$. Single-mode Gaussian gates include phase shifts, displacement, and squeezing. Two-mode gates include beamsplitters. A combination of these gates can be used to implement gates such as quadratic phase gates, controlled-phase gates, Mach-Zehender interferometers, etc. On the other hand, non-Gaussian operations do not preserve the Gaussian nature of the quadratic states. These include the cubic phase gate, the Kerr gate, and the cross-Kerr gate. Expressions of all these gates are given in Table~\ref{tab:gates}. It has been shown that all these Gaussian transformations combined with any single non-Gaussian transformation make a universal gate set for PQC~\cite{PhysRevLett.82.1784}. 

\begin{table}[t]
\caption{Currently available qumode operations as gates. In each gate, the argument shows adjustable parameters. $i$ and $j$ denote two arbitrary qumodes.}
\label{tab:gates}
\begin{tabularx}{\columnwidth}{>{\raggedright\arraybackslash}p{0.25\columnwidth}>{\raggedright\arraybackslash}X>{\raggedright\arraybackslash}p{0.25\columnwidth}}
\toprule
\textbf{Gate} & \textbf{Expression} & \textbf{Type} \\
\midrule
Rotation or Phase-shift & $R(\phi) = \exp[i\phi \hat{n}]$ & Gaussian \\
Displacement & $D(\alpha) = \exp[\alpha\hat{a} - \alpha^*\hat{a}^\dagger]$ & Gaussian \\
Squeezing & $S(r, \phi) = \exp\left[\frac{r}{2}(e^{-i\phi}\hat{a}^2 - e^{i\phi}\hat{a}^{\dagger 2})\right]$ & Gaussian \\
Beamsplitter & $BS(\theta,\phi)= \exp \left[ \theta \left( e^{i\phi} \hat{a}_i \hat{a}^\dagger_j - e^{-i\phi} \hat{a}^\dagger_i \hat{a}_j\right) \right] $ & Gaussian  \\
\midrule
Quadratic Phase & $P(s) = \exp\left[i\frac{s}{2 \hbar} x^2\right]$ & Gaussian (Decomposable) \\
Controlled-Phase & $CZ(s)= \exp\left[ i s \frac{\hat{x}_i\hat{x}_j}{\hbar}\right]$ & Gaussian (Decomposable) \\
Two-mode squeezing & $S_2(z)= \exp \left[  \ z \hat{a}^\dagger_i \hat{a}^\dagger_j - z^*\hat{a}_i \hat{a}_j \right] $ & Gaussian (Decomposable) \\
\midrule
Cubic Phase & $V(\gamma)= \exp\left[i \frac{\gamma}{3\hbar}\hat{x}^3\right]$ & Non-Gaussian \\
Kerr & $K(\kappa)= \exp\left[i\kappa\hat{n}^2\right]$ & Non-Gaussian \\
Cross-Kerr & $CK(\kappa)=\exp\left[i\kappa\hat{n}_i\hat{n}_j\right]$ & Non-Gaussian \\
\bottomrule
\end{tabularx}
\end{table}

In the decoding stage, the measurement scheme can be homodyne, heterodyne, or photon number-resolving measurements. The choice of measurement scheme is determined by the nature of the encoding and the specific variables of interest, either position or number operators.

Based on the gates available in qumode-based architectures, one of the challenges in implementing problem-inspired algorithms is to implement problems represented by polynomials of degree $d\geq 3$ as all the Gaussian operations are quadratic in terms of phase-space quadratures. Even for $d=3$, implementing a cubic-phase gate is required, which is a non-Gaussian operation. For $d>3$, we have to decompose the operations with higher degrees in terms of available gates~\cite{PhysRevA.99.022341}. This challenge applies to both phase space and Hilbert space encodings. Consequently, there is a need for an algorithm that can efficiently handle functions of any degree using a low number of gates, making it feasible for implementation on currently available photonic devices. In the following sections, we explore how counterdiabatic protocols can address this requirement and enable the efficient implementation of high-degree functions in PQC. We accomplish this by providing examples illustrating how both formulations can be used to encode different types of problems and demonstrate the effectiveness of our algorithm in both cases.

\section{Photonic counterdiabatic quantum optimization} \label{sec:counterdiabatic}
\begin{figure*}
    \centering
    \includegraphics[width=1\linewidth]{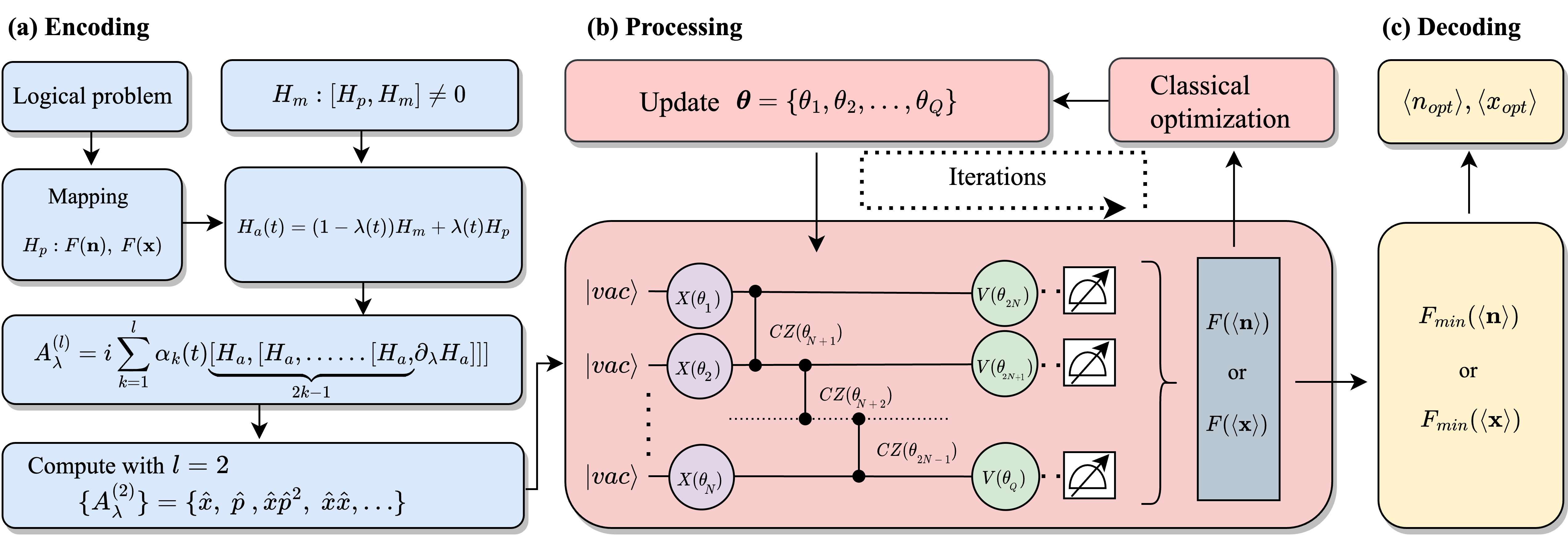}
    \caption{A schematic diagram illustrating the $p=1$ photonic counterdiabatic quantum optimization (PCQO) algorithm. (a) Encoding phase: A logical problem, denoted as $H_p$, is encoded into $F(\mathbf{n})$ or $F(\mathbf{x})$ based on the problem type. The mixer $H_m$ is introduced satisfying the non-commutativity relation, enabling the generation of the operator pool using the nested commutator method. (b) Processing phase: The operators from $\mathcal{A}=\{A_{\lambda}^{(2)}\}$ are exponentiated and employed as a circuit ansatz $U_{cd}(\boldsymbol{\theta})$ with adjustable parameters  $\boldsymbol{\theta}=\{\theta_1, \theta_2,\ldots, \theta_Q\}$.  $Q$ shows the total number of parameters. The algorithm initiates with random parameter values and iteratively updates them through classical optimization, aiming to determine $F(\braket{\mathbf{n}})$ or $F(\braket{\mathbf{x}})$ until convergence is achieved. (c) Decoding phase: Performing measurements and extracting solutions from the minimum values, $F_{min}(\braket{\mathbf{n}})$ or $F_{min}(\braket{\mathbf{x}})$, enables the representation of solutions in the form of the mean photon number $\braket{\mathbf{n}}$ or the mean quadrature values $\braket{\mathbf{x}}$.  }
    \label{fig:schematic}
\end{figure*}

To design the algorithm, we start with an adiabatic quantum Hamiltonian $H_a(t)$, given by
\begin{equation}
   H_a(t) = (1 - \lambda(t))H_m + \lambda(t)H_p \label{aevol}
\end{equation}
where $\lambda(t)$ is a scheduling function such that $\lambda(0)=0$ and $\lambda(T)=1$ and $T$ is the total evolution time. $H_p$ is a problem Hamiltonian whose ground state we need to find and $H_m$ is a mixer Hamiltonian, such that $[H_p,H_m]\neq0$. According to the adiabatic theorem, if the system is prepared in an eigenstate of $H_m$, it remains in the instantaneous eigenstate during the evolution, provided that the evolution is slow such that  $|~\dot{\lambda}~| \ll 1$. A universal quantum computing paradigm, called adiabatic quantum computing, was developed based on this Hamiltonian~\cite{RevModPhys.90.015002}. This paradigm was extended to digital quantum computing with experiments on a superconducting circuit using Trotterization~\cite{Barends2016}. One of the VQAs that take inspiration from this is QAOA. In QAOA, two non-commuting unitaries called the mixer term $U_b(\beta)$ and the Hamiltonian term $U_c(\gamma)$ are applied iteratively for $p$ layers to an initial state $\ket{\psi_0}$, which is the ground state of $H_m$. $\gamma$ and $\beta$ are optimizable parameters, and the parameterized unitary looks like
\begin{equation}
    U(\vec{\gamma}, \vec{\beta}) = \prod_{k=1}^p U_b(\beta_k)U_c(\gamma_k). \label{qaoa_form}
\end{equation}
where, $U_b(\beta)=e^{-i\beta H_m}$, $U_c(\gamma)=e^{-i\gamma H_p}$. The mixer Hamiltonian $H_m$ can take various forms depending on the specific requirements of the problem being solved \cite{a12020034}.
The difficulty that comes with adiabatic algorithms is the requirement of deep circuits to satisfy the slow evolution condition. If the evolution is not sufficiently slow, there will be diabatic transitions that will reduce the probability of finding the system in the ground state of $H_p$. In order to get fast evolutions, CD protocols are implemented. Here, the task is to add velocity-dependent terms $A_{\lambda}^{(l)}$ to minimize the non-adiabatic transitions. This results in the Hamiltonian $H_{cd}$ given by
\begin{equation}
H_{cd}(t) = H_a(t) + \dot{\lambda}(t) A_{\lambda}^{(l)}. \label{cdevol}
\end{equation}
The calculation of the exact CD term requires full spectral information~\cite{Demirplak2003,Berry_2009}. This information may not always be available hence approximate CD terms can be used instead~\cite{doi:10.1073/pnas.1619826114}. One of the ways to obtain these terms is to utilize adiabatic gauge potentials that can be calculated using the nested commutator method~\cite{PhysRevLett.123.090602} given by
\begin{equation}
    A_{\lambda}^{(l)} = i \sum_{k = 1}^l \alpha_k(t) \underbrace{[H_{a},[H_{a},\dots,[H_{a},}_{2k-1}\partial_{\lambda} H_{a}]]],
    \label{gauge}
\end{equation}
where $l$ is the order of expansion and $\alpha_k(t)$ are the CD coefficients that need to be optimized. This can be done analytically~\cite{PhysRevLett.123.090602} or by using variational circuits~\cite{doi:10.1098/rsta.2021.0282}.

Now, let us assume that there exists a $\lambda$ such that $|~\dot\lambda~| \gg 1$ at the beginning of the evolution. For this condition, the $H_a$ term from Eq.~\eqref{cdevol} can be neglected because only contributions that will dominate will be from the CD terms $A_{\lambda}^{(l)}$. Thus, we can devise a parameterized unitary evolution that will look like
\begin{equation}
U(\boldsymbol{\theta}) =\prod_{k=1}^p U_{cd}(\boldsymbol{\theta}_k),
\end{equation}
where $U_{cd}(\boldsymbol{\theta})=e^{-i \boldsymbol{\theta} \mathcal{A}}$. $\mathcal{A}$ is a set of operators obtained by Eq.~\eqref{gauge} with low order of expansion and $\boldsymbol{\theta}_k$ is a set of tunable parameters. This method has been shown to reduce the circuit depths drastically as compared to other adiabatic algorithms in qubit-based technologies~\cite{chandarana2022digitized}. 

The adiabatic theorem is constructed under the assumption that both $H_m$ and $H_p$ have discrete spectra and there exists a finite minimum spectral energy gap for all times $t$~\cite{farhi2000quantum}. However, in the CV regime, this is not always guaranteed since the evolution cannot be bounded with a well-defined spectral gap. This motivated proposals for equivalent adiabatic theorem for continuous systems~\cite{PhysRevLett.101.150407}. In any case, we can always discretize the time and implement the evolution in terms of a gate-based model. The main motivation for this work is to develop hybrid CD-inspired protocols for CV systems, specifically PQC. The expected advantages of doing this are two-fold. Firstly, these methods should result in a reduction in the required number of operations. Since the operations in photonics are beam-splitters and interferometers which are imperfect, this reduction should improve the performance to a great extent. Secondly, since at a finite order $l$, we get a pool of CD operators from which we can choose suitable operations that are natively available in the device. This makes the algorithm more flexible for NISQ devices. As we will see later, this freedom will also allow us to attempt complicated problems without the need for decompositions. This is crucial since decompositions are non-trivial and require lots of resources in the PQC regime.

Now, we devise a CD-inspired algorithm for PQC. To do so, we start by considering a problem Hamiltonian $H_p$. This can be a function of $\mathbf{x} = (x_{1},\ldots ,x_{N}) \in \mathbb{R}^{N}$ or a function of $\mathbf{n} = (n_{1},\ldots ,n_{N}) \in \mathbb{W}^{N}$ depending upon the encoding. Then we define a mixer Hamiltonian $H_m$ whose ground-state is easy to prepare and satisfies the condition $[H_p, H_m]\neq 0$. With $H_m$ and $H_p$, we now have $H_a$ given by Eq.~\eqref{aevol}. From this, we can compute $A_{\lambda}^{(l)}$ using Eq.~\eqref{gauge} with a specific order $l$ to get a pool of operators. We heuristically select operators from this pool and parameterize them. This selection is made based on the requirements of the problem and the hardware. Starting from a specific initial state, these parameters can be optimized by classical optimization routines to minimize $F(\braket{\mathbf{x}})$ or $F(\braket{\mathbf{n}})$. A schematic diagram of the algorithm is shown in Fig.~\ref{fig:schematic}. 

The cost function used in this algorithm focuses on minimizing the mean values of the operators $\braket{\hat{x}}$ or $\braket{\hat{n}}$ instead of optimizing the expectation value $\braket{F(\mathbf{x})}$ or $\braket{F(\mathbf{n})}$. This provides a useful approach for hybrid algorithms seeking approximate solutions. By considering the mean values along with the optimal state, the cost function takes into account the probability of obtaining near-optimal states. Moreover, this cost function can exhibit robustness against noise since it relies on the mean of local operators' expectation values rather than the specific state itself. This robustness can be advantageous in practical implementations where noise and imperfections are inevitable. For example, if we want to solve $F(n)= (n-1)^2$, the circuit ansatz will optimize the parameters such that $\braket{\hat{n}}=1$. Hence, along with $\ket{n=1}$, $\ket{n=0}$ and $\ket{n=2}$ will also have finite probabilities. 

\section{Results: Phase-space encoding}\label{sec:phase_space}
We examine the performance of the algorithm by applying it to two non-convex classical optimization problems $F(\mathbf{x})$, both represented as polynomials of degree $d$. For both problems, the initial state preparation involves setting all the qumodes to vacuum states. The mixer Hamiltonian $H_m = \sum_i(\hat{p_i}-p_0)^2$ was selected, where $p_0$ is a constant. This resembles the kinetic part of the harmonic oscillator and the potential part would be $F(\mathbf{x})$.  This mixer satisfies the non-commutativity condition with $F(\mathbf{x})$. The next step involves finding the operator pool $\mathcal{A}$ by evaluating $A_{\lambda}^{(l)}$ which enables us to choose suitable parameterized gates from $\mathcal{A}$ as a circuit ansatz. We utilize homodyne measurements to compute $\braket{\mathbf{x}}$. Since this is a numerical analysis, we have to define a cutoff dimension of the Fock space which was set to $D=15$ for the toy function, $D=10$ for the Rosenbrock function due to computational limitations and we also set $\hbar=2$. We considered two polynomials, one with $d=4$ and $N=4$ variables and the other with $d=6$ and $N=3$ variables. 

We start with the Rosenbrock function~\cite{Rosenbrock1960}, defined as
\be
F(\mathbf{x}) = \sum_{i=1}^{N-1} \left[100(x_{i+1}-x_{i}^{2})^{2} + (1-x_{i})^{2}\right] 
\ee
where $\mathbf{x} = (x_{1},\ldots ,x_{N}) \in \mathbb{R}^{N}$. Due to its non-convex nature and difficulty in reaching global optima, this function is widely used as a benchmark for optimization algorithms. We consider $N=4$ case with global minimum $F_{min}(\mathbf{x}_{opt})= 0 $ at $\mathbf{x}_{opt}= (1,1,1,1)$. We limit ourselves to the $A_{\lambda}^{(2)}$ and get $\mathcal{A} = \{\hat{p},~\hat{x}\hat{p}+\hat{p}\hat{x},~\hat{x}^3,\ldots\} $. This will contain many higher-order terms but we heuristically select the first three terms. These correspond to the $X$ gate, $S_2$ gate, and $V$ gate (See Table~\ref{tab:gates}). This choice was made to include single-mode Gaussian, two-mode Gaussian, and non-Gaussian gates. Since we will perform Gaussian measurements, these non-Gaussian gates will make sure that the ansatz is not effectively simulated classically~\cite{PhysRevLett.88.097904}.  Since $S_2$ is a two-mode gate, we have to decide the connectivity. For simplicity, we keep this connectivity to the nearest neighbors but this can be further fine-tuned based on hardware constraints. Therefore,  the circuit ansatz will have $X$ gates applied to all qumodes, $S_2$ gates applied to nearest-neighbor qumodes, and $V$ gates applied to all qumodes and thus the number of parameters required will be $Q = 2N + (N-1) = 3N-1$.

\begin{figure}
    \centering
    \includegraphics[width=1\linewidth]{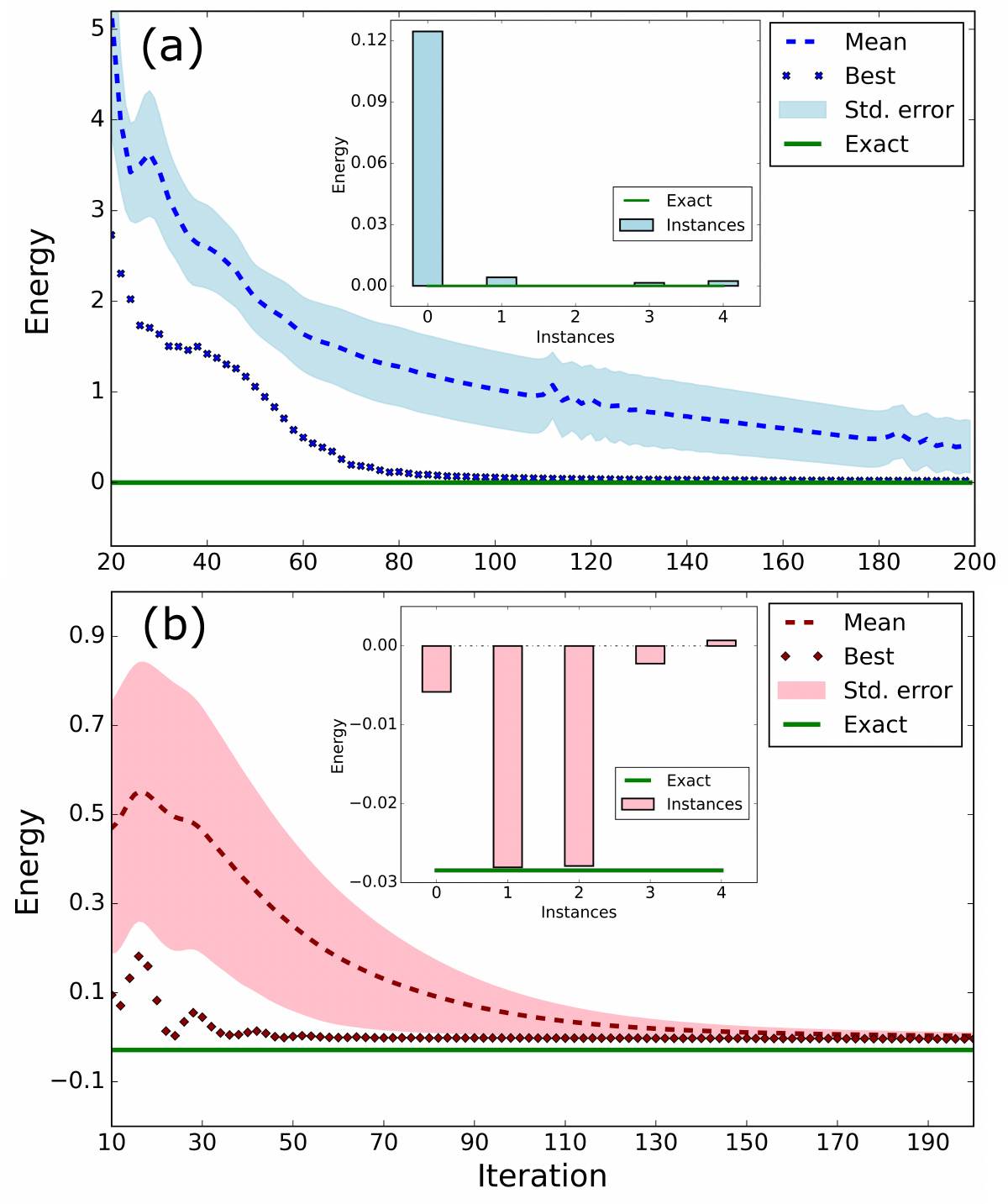}
    \caption{Energy (values of $F(\braket{\mathbf{x}})$) profiles for (a) the Rosenbrock function and (b) the toy function (Eq.~\ref{eq:toyfunction}) with a $p=1$ ansatz. Dashed lines represent the mean of five random initializations, solid lines indicate the best outcome among the five instances, and shaded regions show the standard error. The green line represents the exact energy. The inset plots display the minimum energies achieved with the five instances.}
    \label{fig:classicalfunctions}
\end{figure}

The second benchmark function is a toy problem given by
\begin{equation}\label{eq:toyfunction}
\begin{aligned}
F(\mathbf{x}) =\ &(x_1^3 + x_2^3 + x_3^3 - 1x_1 + 2x_2 - 3x_3)^2 +\\
&(-x_1 + x_2 + x_3)^2 + 0.01x_1 + 0.01x_2 + 0.01x_3.
\end{aligned}
\end{equation}
For this problem, the solution is $F_{min}(\mathbf{x}_{opt})= -0.028457$ at $\mathbf{x}_{opt}= (-1.42212,-0.127017,-1.29723)$. This function is selected because its degree is much larger than the degree of the operations implementable in current technologies (See Table.~\ref{tab:gates}). Furthermore, this function forces correlations between the variables, so the problem cannot be solved independently for each variable. The linear terms with low coefficients ensure that the global minima are reached only at a single point. This function will give the same lower-order operators in the $\mathcal{A}$ pool as the Rosenbrock function hence we implement the same circuit ansatz as before. In both cases, the mean values of $F(\braket{\mathbf{x}})$ (termed as energy) over five random initializations across several iterations for $p=1$ layer of the ansatz are computed and results are shown in Fig.~\ref{fig:classicalfunctions} along with the standard error. Energy variation corresponding to the best instance is also plotted. For parameter optimization, we have used the Adam optimizer~\cite{kingma2014adam}. 

Fig.~\ref{fig:classicalfunctions}(a) illustrates the convergence of the mean energy over the first 200 iteration steps out of 500, for the Rosenbrock function, with some iterations skipped to disregard initial fluctuations. It can be observed that the mean energy is slightly higher than the exact energy required to solve the problem. This indicates that the algorithm's performance is affected by the initial parameters chosen. This observation is supported by the fact that the standard error is high in later iterations. In the best-instance run, the algorithm achieves the exact solution. The inset shows the minimum energy achieved during the optimization with the five instances. We can see that, for most instances, the algorithm results in solutions very close to the exact solution.  This demonstrates that if the objective is to obtain approximate solutions, the algorithm exhibits good performance. 

Fig.~\ref{fig:classicalfunctions}(b) displays the convergence of the mean energy over the first 200 iterations out of 1000, for the toy function in Eq.~\ref{eq:toyfunction}. It can be observed that the mean energy converges close to the exact energy within the first 200 iterations. Initially, there is a high standard error, but in later iterations, the algorithm successfully finds approximate solutions. At the 200th iteration, the best instance achieves $F(\mathbf{x})=-0.0037$ while the mean energy is $F(\mathbf{x})=0.0029$. Both values are sufficiently close to the exact solution. The inset plot shows the minimum energy obtained during the 1000 iterations for all instances. The best solution we get is $F(\mathbf{x})=-0.0280$. 

For both cases, we noticed that optimization becomes slower after about 200 iterations. Due to this, a larger number of iterations to achieve more accurate solutions is required. A possible explanation for this is that the ansatz reaches a point where the energy landscape becomes flat which results in low gradient values. Consequently, more iterations are needed to reach the solution. This issue can potentially be addressed by adding more terms to the ansatz, although this may introduce a more rugged solution landscape with potential local minima. Nonetheless, for obtaining approximate solutions, the algorithm performs well even with random initializations. 

The results demonstrate that the proposed PCQO algorithm can obtain good approximate solutions with just $p=1$ layer of the ansatz. This algorithm can be extended to handle polynomials of any degree, and it would be intriguing to explore its performance on higher-degree polynomials. Additionally, investigating problems with a larger number of variables would provide further insights into the algorithm's capabilities and potential applications. 

To tackle these problems with qubit-based algorithms, the system size will depend upon the bit resolution that needs to be achieved. This will lead to large resource requirements and the cost functions will be many-body Ising Hamiltonians~\cite{stein2023evidence}. Since the requirement to discretize the solution here is elevated, any arbitrary precision can be reached using the PCQO algorithm with a linear encoding with respect to the variables. These advantages make PCQO far more suitable for tackling continuous optimization problems. 

Despite the advantages, some challenges need to be addressed. For instance, clever optimization techniques need to be developed to find the optimal parameters faster. Strategies to efficiently choose the circuit ansatz from $\mathcal{A}$ ensuring trainability and expressibility need to be developed since there is no performance guarantee of the heuristic choice of ansatz. Nevertheless, PCQO works extremely well for phase-space encoding. In the next section, we will investigate the performance of the algorithm using Hilbert-space encoding where the states are countably infinite and discrete.

\section{Results: Hilbert-space encoding}\label{sec:hilbert_space}
In Hilbert-space encoding, the variables are represented by Fock states $\ket{n}$. This encoding scheme naturally lends itself to tackling integer programming problems, which are a class of optimization problems where some or all variables are constrained to be integers. Integer programming can be further categorized into three subclasses: linear integer programming, non-linear integer programming, and mixed-integer programming. The classification depends on the linearity of the cost function and constraints. Here, we focus on solving linear integer and non-linear programming problems, but the same techniques can be readily extended to mixed-integer programming as well. 

Similar to the phase-space encoding, the initial state of all the qumodes is prepared in the vacuum state. The mixer we selected was $H_m = \sum_i~(x_i-x_0)^2 + (p_i-p_0)^2$ where $x_0$ and $p_0$ are constants. The mixer resembles a shifted harmonic oscillator in both quadratures. Instead of performing homodyne measurements, number-resolving measurements are used to determine the values of $\braket{\mathbf{n}}$. These measurements provide information about the number of photons in each mode, which is essential in the Hilbert space encoding.
\subsection{Unbounded knapsack problem}
\begin{figure}
    \centering
    \includegraphics[width=1\linewidth]{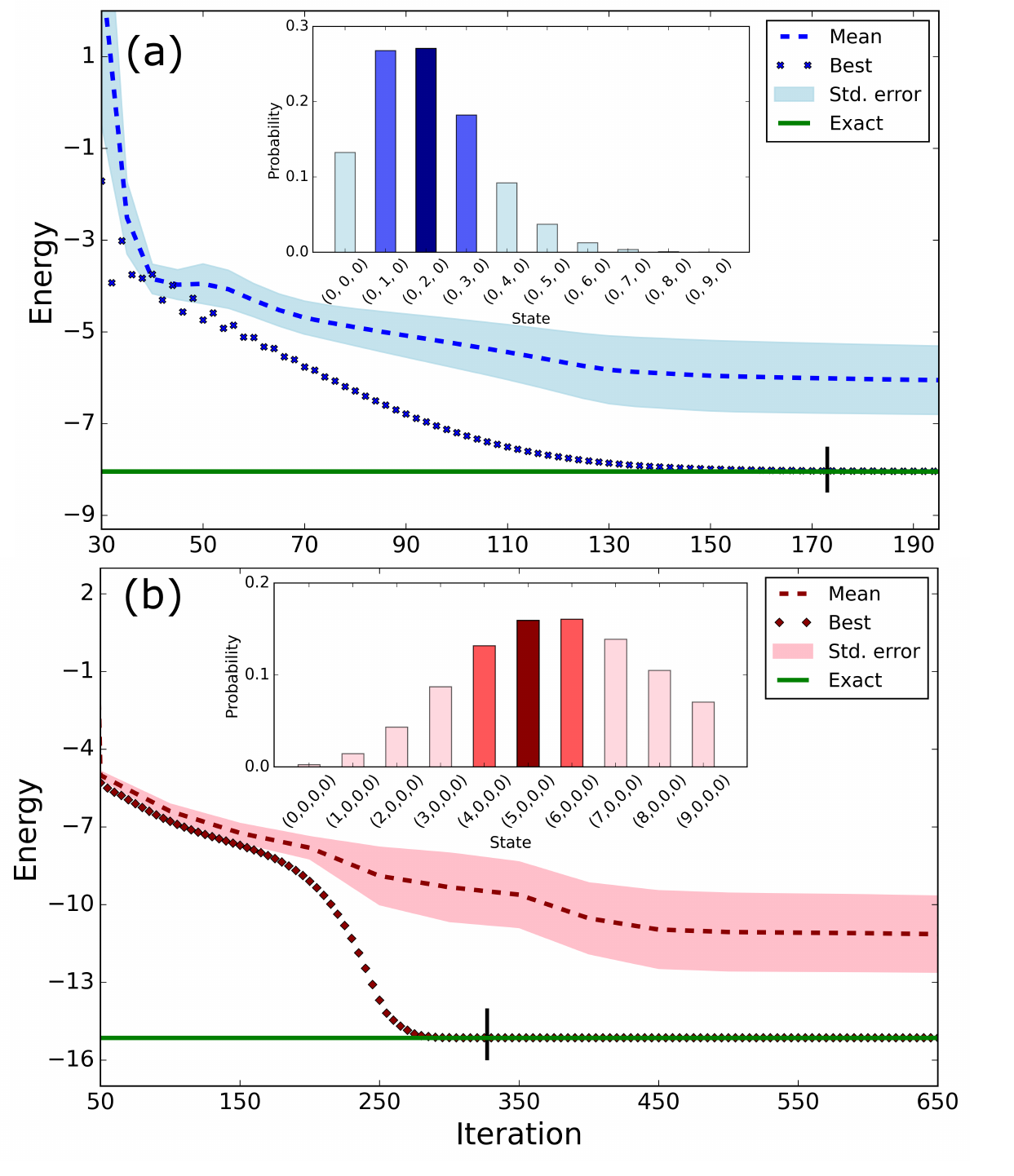}
    \caption{Energy evolution over iteration steps for (a) $N=3$ and (b) $N=4$ qumode knapsack problems using a layered ansatz with $p=1$. Dashed lines represent the mean energy of 5 random initializations, while shaded regions indicate the corresponding standard error. The solid lines depict the best energy obtained among the 5 random instances. The green lines indicate the exact energy minimization. Vertical black lines denote the iteration at which $10^{-2}$ accuracy relative to the exact energy was reached during optimization. Insets show the probability distribution of Fock states around the solution up to the cutoff.  }
    \label{fig:ukp}
\end{figure}

We commence with investigating a small instance of the unbounded knapsack problem (UKP)~\cite{Lueker1975}. In UKP, we consider a set of different types of items $i$, each with a value $v_i$ and weight $w_i$. The objective is to maximize the total value while ensuring that the total weight of the selected items does not exceed the capacity $C$ of the knapsack. Unlike the bounded knapsack problem, the UKP allows for an unlimited number of items of the same type to be included in the knapsack. If the number of items of type $i$ that can be included in the knapsack is given by $n_i$, then the optimization problem looks like
\be
\begin{aligned}
\min_{n_i}\ -\sum_{i=1}^N v_in_i \quad \text{subject to} \quad \sum_{i=1}^N w_in_i \leq C,\ n_i \geq 0 \ \forall i,
\end{aligned} \label{ukpprob}
\ee
where $N$ shows the total items. This problem is classified as an integer linear programming problem, where both the cost function and constraints are linear functions of $n_i$. This can be converted into a minimization problem of
\be
F(\mathbf{n})= -\sum_{i=1}^N v_in_i + \delta \sum_{i=1}^N ( w_i n_i - C)^2. \label{ukpfn}
\ee
Here, $\delta$ represents the penalty term. Usually, an auxiliary variable is added to account for the inequality. However, as this algorithm is aimed at finding approximate solutions, we add the inequality as a `soft' constraint in the problem Hamiltonian. We utilized the $p=1$ layered PCQO algorithm to address the UKP for $N=3$ and $N=4$ qumodes. 
\begin{table}[t]
\caption{The data considered for the UKP instances where $N$ are the total items,  $v_i$ is the value of the item,  $w_i$ is the weight of the item, $C$ is the total capacity of the knapsack, $F_{min}(\mathbf{n}_{opt})$ shows the optimal cost function at $\mathbf{n}_{opt}$ optimal values.  }
\begin{ruledtabular}
\begin{tabular}{cccccc}\label{ukpdata}
$N$ & $v$ & $w$ & $C$ & $F_{min}(\mathbf{n}_{opt})$ & $\mathbf{n}_{opt}$ \\
\colrule
$3$ & $[3,4,1]$ & $[9,5,8]$ & $10$ & $-8$ & $(0,2,0)$ \\
$4$ & $[3,4,1,3]$ & $[2,7,6,6]$ & $10$ & $-15$ & $(5,0,0,0)$ \\
\end{tabular}
\end{ruledtabular}
\end{table}

We considered $l=2$ order nested commutator, which will result in $\mathcal{A} = \{\hat{x},~ \hat{p},~ \hat{x}\hat{p}^2,~ \hat{x}\hat{x},\ldots \}$. From this, we choose $\hat{x}$ and $\hat{x}\hat{x}$, whose exponentiation will correspond to $X$ gates and $CZ$ gates respectively. We do not include the non-Gaussian gates because the measurement is non-Gaussian, which will restrict the effective classical simulation~\cite{PhysRevLett.88.097904}. Thus the circuit ansatz consists of $X$ gates applied to all qumodes and $CZ$ gates applied to nearest neighbor connections. Hence, the number of parameters required will be $Q= N + (N-1) = 2N-1$. The cutoff dimension was chosen as $D=10$, the penalty term was set to $\delta=4$, and we keep $\hbar=2$. Mean energy (values of $F(\braket{\mathbf{n}})$) and the best instance as a function of the number of iterations for both $N=3$ and $N=4$ qumode case are shown in Fig.~\ref{fig:ukp}. In both cases, we skip some iterations to neglect the initial fluctuations. As before, classical optimization was performed by using the Adam optimizer. The data considered for the problems are given in Table.~\ref{ukpdata}.

Fig.~\ref{fig:ukp}(a) depicts the energy over 200 iterations for the $N=3$ UKP. The mean energy converges to a value higher than the exact solution. This is due to initialization, which can cause solutions to fall into local minima. This claim is supported by the high standard error even in later iterations. Efficient initialization strategies are crucial for mitigating the dependence on initial points in VQAs. However, the best instance achieves the exact energy. The black vertical line represents the instance where $10^{-2}$ accuracy was first attained, around the 170th iteration. This indicates that with optimal initial parameterization, the algorithm performs exceptionally well and reaches the exact energy. Fig.~\ref{fig:ukp}(b) presents the mean energy variation across 700 iterations for the $N=4$ UKP. Same as before, we observe that the mean energy is higher than the exact energy, and the best case reaches the exact energy. However, in contrast to the previous case, it takes 350 iterations to achieve a tolerance of $10^{-2}$, which is significantly higher.

It is important to note a subtle caveat in our approach. As we compute $\langle F(\mathbf{n})\rangle$ as the cost function, the solution space is continuous because $\langle \hat{n}\rangle$ can take any non-negative values. Consequently, post-processing techniques are necessary to extract the solution. This can be accomplished in two ways. Approximating the solution to the nearest integer or analyzing the probability distribution of the Fock states to determine the likelihood of obtaining the desired state. The inset plots in the figure display the probability distribution of states near the optimal solution for the best instance. In the case of $N=3$, where the optimal solution is represented by $\mathbf{n}_{opt}=(0,2,0)$, the probabilities were plotted for states where the 2nd qumode varies from $\ket{n_1}=0$ to $\ket{n_1}=D-1$. Similarly, for the $N=4$ case, with the optimal solution of $\mathbf{n}_{opt}=(5,0,0,0)$. The darkest color in the plots indicates states that achieve high probabilities, representing the optimal solutions. However, we can also observe that approximate solutions have relatively high probabilities also, depicted by lighter colors. This occurrence arises because our cost function optimizes the mean, allowing sub-optimal states to have finite probabilities. This characteristic is crucial for approximate optimization algorithms since their objective is to find approximate solutions rather than exact ones. For the $N=3$ case, the sum of probabilities for states $\ket{n} = (0,1,0)$, $\ket{n}=(0,2,0)$, and $\ket{n}=(0,3,0)$ amounts to approximately 72\%. Similarly, in the $N=4$ case, the sum of probabilities for states $\ket{n}=(4,0,0,0)$, $\ket{n}=(5,0,0,0)$, and $\ket{n}=(6,0,0,0)$ totals around 45\%. So, even when the mean photon number is minimized, we can extract the solution successfully. 

It is straightforward to notice that there is an improvement in terms of the quantum resources required to encode a problem using qubits. For encoding all possible solutions to this problem, we need the number of qubits $N_{\text{qubits}} = \sum_{i=1}^N \bigg\lceil \log_2 \left( \bigg\lfloor \frac{C}{w_i} \bigg\rfloor \right) \bigg\rceil \geq N$. This scales worse than linearly with the number of items. $N_{\text{qubits}}$ corresponds to the number of qubits necessary to encode trivial solutions in which we only include a single item until we fill the knapsack, giving an idea of the order of the resources required. On the other hand, when employing our approach, we can trivially see that the number of qumodes scales linearly with the number of items.

In conclusion, the PCQO algorithm works considerably well for the UKP problem cases we considered. The energy profile motivates the investigation of techniques that find suitable initial parameters to increase the performance of the algorithm. In QAOA-like algorithms, we have to implement non-Gaussian operations in the ansatz whereas PCQO can perform the optimization using only Gaussian operations. The only non-Gaussian element that is introduced is the number-resolving measurement. As the non-Gaussian operations are relatively hard to realize experimentally, the PCQO algorithm becomes a preferable candidate for near-term photonic devices solving integer programming problems.
\subsection{Maxclique problem}
\begin{figure}
    \centering
    \includegraphics[width=1\linewidth]{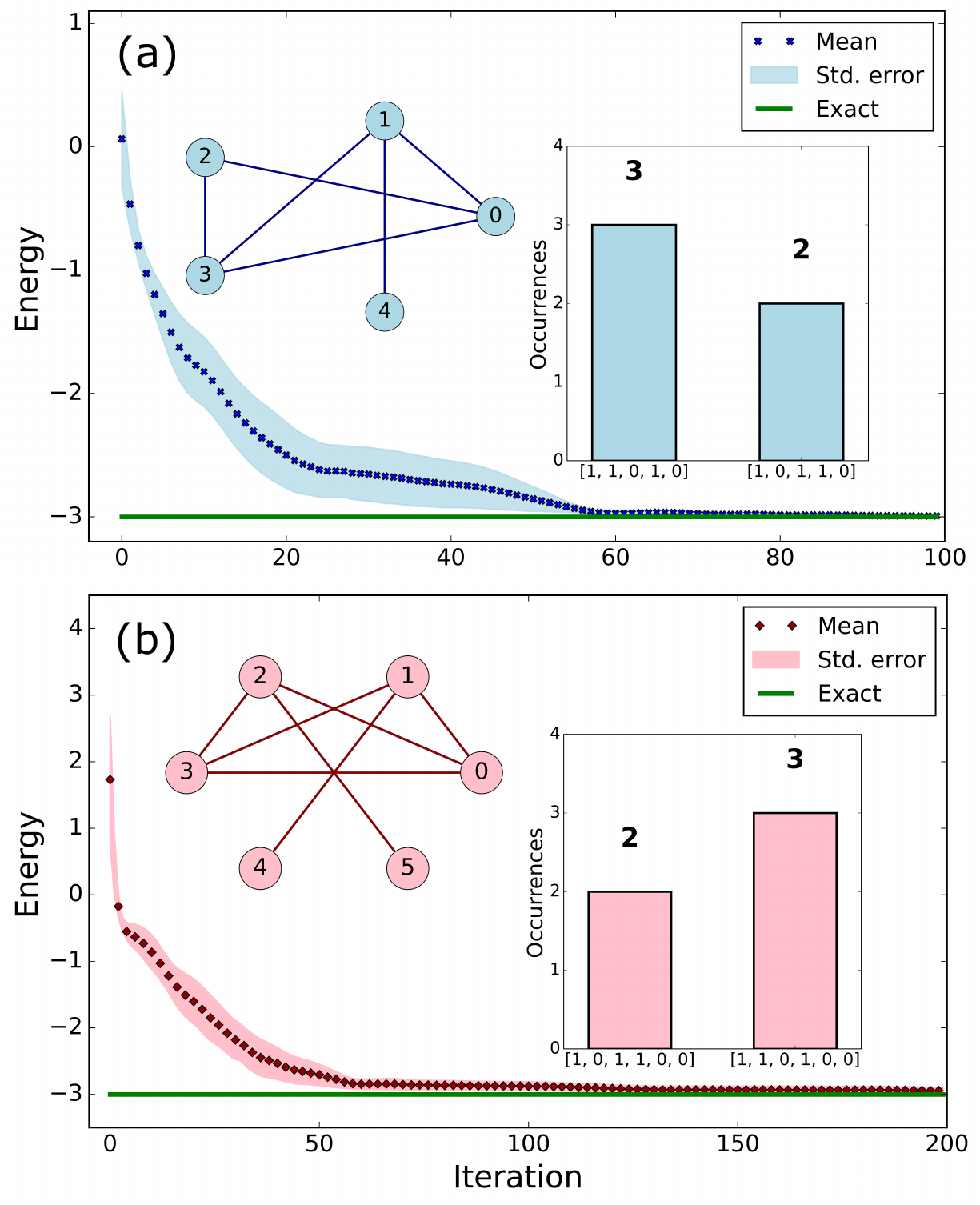}
    \caption{Energy as a function of iteration steps for the $N=5$ and $N=6$ qumode Maxclique problems with a $p=1$ layer ansatz. Solid lines denote the mean energy of 5 random initializations, while shaded regions represent the standard error. (a) $N=5$ qumode Maxclique problem, (b) $N=6$ qumode Maxclique problem. Insets show corresponding graphs and bars for occurrences of the degenerate solutions. Green lines show the exact energy. }
    \label{fig:maxclique}
\end{figure}
The final benchmark we consider is the Maxclique problem~\cite{Bomze1999}. Consider a graph $\mathcal{G}=(V, E)$ with $V$ vertices and $E$ edges that can be represented by its adjacency matrix $Z$, where $Z_{ij}=1$ if there exists an edge between vertices $i$ and $j$, and $Z_{ij}=0$ otherwise. The Maxclique problem involves finding the largest subset of vertices $V_s \subseteq V$ in graph $\mathcal{G}$, where all vertices in $V_s$ are mutually connected.

To encode the Maxclique problem in terms of Fock states, we introduce binary variables $n_i \in \{0, 1\}$, representing the selection of a vertex $i$ in the maximum clique. This leads to 
\be
\begin{aligned}
\min_{n_i}\ -\sum_{i \in V} n_i\quad \text{subject to}\quad \sum_{i,j \in V} (\mathbf{I}-Z_{ij})~n_in_j = 0,
\end{aligned} \label{maxcliqrob}
\ee
where $\mathbf{I}$ is an identity matrix. Here, the objective is to maximize the sum of selected vertices $n_i$, indicating the size of the maximum clique. The constraint ensures that if vertices $i$ and $j$ are not connected by an edge, they cannot both be selected in the maximum clique. Hence the problem Hamiltonian in terms of number operators will be
\begin{equation}
   F(\mathbf{n}) = -\sum_{i \in V} n_i + \delta_1 \sum_{i,j \in V} (\mathbf{I}-Z_{ij})~n_in_j + \delta_2 \sum_{i \in V} n_i(n_i-1)
\end{equation}
where the first two terms correspond to the Eq.~\eqref{maxcliqrob} and the third term ensures that the search of the Fock space is restricted to one. Penalty terms $(\delta_1,\delta_2)$ are applied to change the weights of the constraints depending upon the problem instance. These problems come under the class of nonlinear integer programming problems where the cost function and constraints can both be nonlinear functions of integers.  

The PCQO algorithm was applied to find the maximum clique in graphs with $N=5$ and $N=6$ nodes. The mean energy convergence for the $p=1$ layer ansatz is shown in Fig.~\ref{fig:maxclique}(a) for $N=5$ and Fig.~\ref{fig:maxclique}(b) for $N=6$, with the respective graphs shown as insets. The cutoff dimension was $D=5$ due to computational limitations. The penalty terms were set to $\delta_1=10$ and $\delta_2=1$. Adam optimizer was used as a classical optimizer. The maximum cliques obtained were $V_s=\{0,1,3\}$ and $V_s=\{0,2,3\}$ for both $N=5$ and $N=6$ nodes graph. The exact energy was found in both cases regardless of the initial parameters chosen. However, for the $N=5$ case, 200 iterations were needed to reach the exact energy. The inset bar plots show that depending on the initial parameters, the optimal converges to one of the two degenerate solutions for both $N=5$ and $N=6$. It is worth noting that since the Fock states are constrained to be 0 or 1, this algorithm can be implemented with the same resources in qubit-based technologies as well. Therefore, a performance comparison between qumode-based and qubit-based approaches would be interesting.

\section{Comparison with CV-QAOA}
\begin{figure}
    \centering
    \includegraphics[width=1\linewidth]{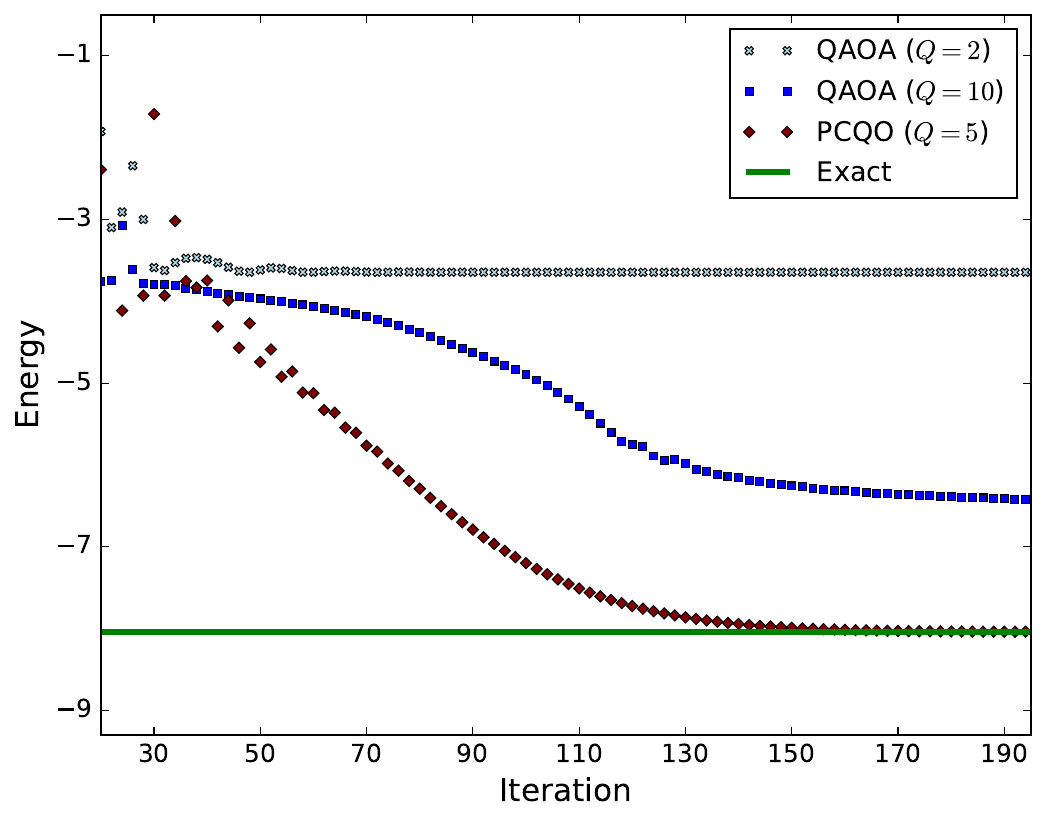}
    \caption{Energy as a function of 200 iteration steps comparing QAOA ($Q=2$, $Q=10$) with PCQO ($Q=5$) for a $N=3$ UKP case. The best instance out of 5 randomly initialized instances is shown. Different markers depict the energy convergence for different algorithms and the green solid line shows the exact energy for the solution.}
    \label{fig:compqaoa}
\end{figure}
In this section, we compare the performance of the PCQO algorithm with two variants of CV-QAOA for a $N=3$ qumode UKP case. In CV-QAOA, we define $H_m= \sum_i p_i^2$ and prepare the initial state as a squeezed state for all qumodes with squeezing parameter $r=1$. The corresponding mixer term is given by $U_{b}(\beta)= \exp(-i\beta \sum_i p_i^2)$, which can be implemented using a custom gate $P_z$ defined as
\begin{equation}
P_z(s) = R\left(- \frac{\pi}{2}\right)P(s)R\left(\frac{\pi}{2}\right),
\end{equation}
where the $R$ gates act as a Fourier transform, rotating the position quadrature into the momentum quadrature. For the Hamiltonian term, we have $U_c(\gamma)= \exp(-i\gamma F(\mathbf{n}))$, where $F(\mathbf{n})$ is determined by Eq.~\eqref{ukpfn}. Implementing this term involves combining $R$ gates, $K$ gates, and $CK$ gates (See Table~\ref{tab:gates}). In PCQO, we initialize with a vacuum state on all qumodes and use the same ansatz as described in the previous section, consisting of $X$ gates followed by nearest-neighbor $CZ$ gates. In both algorithms, we perform number-resolving measurements and optimize the same cost function with cutoff dimension $D=10$.

Regarding parameterization, conventional QAOA has one parameter per unitary, so for the $p=1$ layer, we have $Q=2$. However, in PCQO, we set one parameter per gate, resulting in $Q=5$ parameters for the $p=1$ layer because of the nearest neighbor two-mode gates. To ensure a fair comparison, we also consider a variant of QAOA called MA-QAOA, where each gate has its free parameter~\cite{Herrman2022}. For the $p=1$ layer, this leads to $Q=10$ parameters due to all-to-all connected two-mode gates.
For classical optimization, we have implemented the Adam optimizer for all the cases.

In Fig.~\ref{fig:compqaoa}, we show the energy as a function of iteration steps for the $p=1$ layer in three different algorithms: PCQO and QAOA with $Q=2$ and $Q=10$. The energy values shown correspond to the best outcome out of five randomly initialized instances. We observe that PCQO outperforms both variants of QAOA, achieving the exact energy within 200 iterations. This implies that the operator pool calculated by the nested commutator method serves as a better ansatz for low-layered algorithms compared to the QAOA ansatz. Additionally, the performance of QAOA with $Q=10$ surpasses that of QAOA with $Q=2$ due to the increased degree of freedom given by optimizable parameters. 

As QAOA resembles adiabatic evolution, we might require a high-depth circuit for optimal solutions. It is worth mentioning that the implementation of non-Gaussian gates is approximate at finite cutoff dimensions, and increasing the cutoff could potentially enhance the performance of QAOA in our simulations. Nevertheless, the PCQO algorithm demonstrates superior performance and is particularly suitable for current near-term devices, as it utilizes native Gaussian operations, which are easier to implement experimentally. Regarding phase-space encoding, it becomes apparent that for polynomial functions with degrees higher than three, decomposing them into lower-order gates would be necessary for optimization using the available gates. Such decomposition would require substantial resources that are often unavailable, reinforcing the preference for PCQO as it allows truncating the operator pool to match the available gates. For instance, decomposing $e^{-ix^4}$ requires 29 quadratic gates to be decomposed exactly ~\cite{PhysRevA.99.022341}.  
\begin{figure*}[t]
    \centering
    \includegraphics[width=1\linewidth]{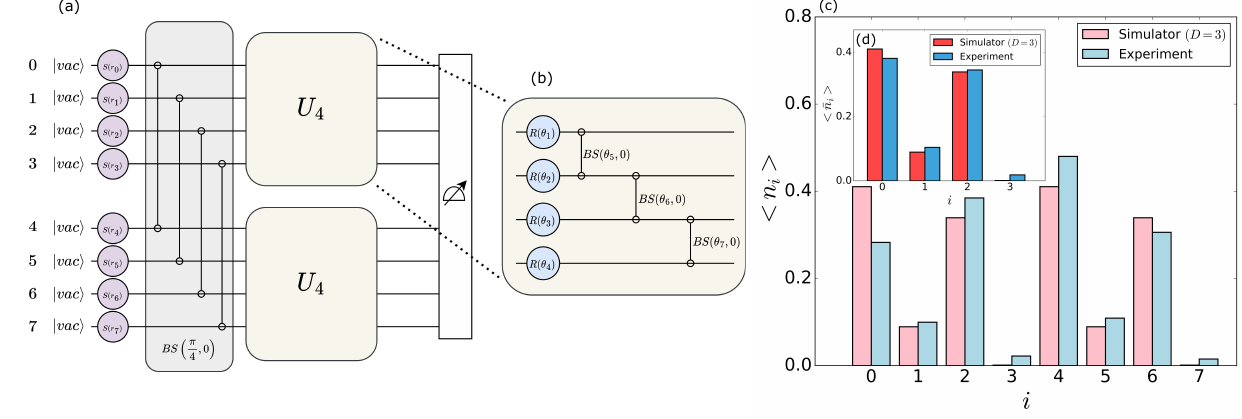}
    \caption{(a) Schematic diagram for eight-qumode nanophotonic chip. The chip is divided into a pair of identical qumodes (0,1,2,3) and (4,5,6,7) utilizing $S_2(r)$ gate. This gate can be decomposed as a $S(r)$ gate ($S(r, \phi=0)$ in Table~\ref{tab:gates}) with $r=0$ or $r=1$ and a $BS\left(\frac{\pi}{4},0\right)$ gate as shown. Then, an arbitrary $U_4$ unitary is applied to the pair followed by number-resolving measurements. (b) shows the PCQO ansatz considered for the experiment. This includes $R(\phi)$ gates applied to all the qumodes and $BS(\theta,0)$ gates applied to nearest-neighbor connections. $\boldsymbol{\theta}= \{\theta_1, \theta_2, \ldots\}$ are optimizable parameters. (c) shows the mean photon number for all the qumodes obtained with the optimal circuit solving  $F(\mathbf{n})=(n_0 + n_2 - 0.75)^2$. The results shown are of a numerical simulation with $D=3$ cutoff and the experiment with 1000 shots. (d) shows the mean photon number obtained by taking an average of the identical qumodes for both numerical simulation and experiment.    }
    \label{fig:experiment}
\end{figure*}
\section{Experimental considerations}\label{sec:experimental}
One of the notable advantages of PQC is its operability at room temperatures~\cite{10.1063/1.4976737}. In the gate-based approach, photonic qumodes are prepared as Gaussian states and manipulated using single or two-mode operations, which can be Gaussian or non-Gaussian~\cite{O'Brien2009,Arrazola2021}. These operations can be implemented using optical waveguides or integrated photonic systems. The PCQO algorithm, designed considering current hardware constraints, follows a specific ansatz. It begins with a vacuum state, applies one and two-mode Gaussian and non-Gaussian operations, and concludes with a homodyne or number-resolving measurement. Implementing single-mode operations such as phase shifts and beamsplitters is straightforward using passive linear optics. Displacement gates can be achieved by utilizing an ancillary qumode~\cite{PhysRevA.71.042308}. However, generating squeezing is challenging as it requires second-order nonlinearity, which can be accomplished using an optical parametric oscillator~\cite{PhysRevLett.57.2520}. Optical parametric oscillators can also generate multimode squeezing and entanglement~\cite{PhysRevLett.68.3663}. Measurement-induced squeezing has also been proposed~\cite{PhysRevA.76.060301}, and experimental implementations of gates like the $CX$ gate have been achieved~\cite{PhysRevLett.101.250501}. Additionally, $CZ$ gates can be implemented experimentally using linear optics and ancillary squeezed vacuum states~\cite{PhysRevA.71.042308}. 

Including a non-Gaussian operation becomes crucial when all other components are Gaussian, as a purely Gaussian state, operations, and measurements can be classically simulated~\cite{PhysRevLett.88.097904}. However, implementation of non-Gaussian gates poses experimental challenges in terms of performance~\cite{PhysRevA.73.062305,PhysRevA.81.043823,PhysRevA.85.033814}. Nevertheless, there are proposals for deterministic cubic-phase gates~\cite{PhysRevLett.124.240503}. Recent proposals have also suggested experimentally feasible continuous-variable quantum neural networks, where non-linearity is achieved through repeat-until-success measurements on ancillary qumodes~\cite{Bangar:22}. 

In terms of measurements, developments have been made in performing homodyne measurements~\cite{Yuen1983,Raffaelli_2018} and number-resolving measurements~\cite{Calkins:13,10.1063/1.5086276}. 
In addition to the circuit ansatz, the classical training part plays a crucial role. To facilitate a gradient-based optimization routine, the gradients need to be obtained from the circuit, which can be achieved using the parameter shift rule~\cite{PhysRevA.99.032331}. Apart from this, recent experiments have demonstrated in situ backpropagation for deep learning~\cite{doi:10.1126/science.ade8450}. Moreover, a quadratic speedup in the optimization of noisy quantum optical circuits has been investigated~\cite{de2023quadratic}. This indicates that the PCQO algorithm can be readily implemented experimentally using currently available devices and can be further extended when large-scale quantum computers with continuous-variable technologies become available~\cite{Fukui_2022}.

To validate the aforementioned claim, we conducted proof-of-principle experiments utilizing a state-of-the-art eight qumode fully-programmable nanophotonic chip~\cite{Arrazola2021}. This 10 mm $\times$ 4 mm chip incorporates two-mode squeezed vacuum states as the initial states, effectively dividing the system into a pair of four independent squeezed vacuum qumodes. The squeezing parameter can be chosen as a binary option of $r=1$ or $r=0$. Thereafter, a programmable SU(4) transformation is applied identically to each of the pair of qumodes. This transformation is based on a network of six Mach-Zehnder interferometers arranged in a rectangular configuration~\cite{Clements:16}. Each of the interferometers has two tunable parameters which give a total of $Q=12$ parameters. Lastly, number-resolving measurements are simultaneously applied to all qumodes. A schematic diagram of the circuit is depicted in Fig.~\ref{fig:experiment}(a).

For the experiment, we considered a two-mode toy problem in the Hilbert-space encoding defined as 
\be
F(\mathbf{n})=(n_0 + n_2 - 0.75)^2.
\ee
For the initial state, the two-mode squeezing parameter was kept to $r=1$ for $(0,4)$ and $(2,6)$ qumodes, and the other two pairs were kept to zero. This was done because the $F(\mathbf{n})$ is a function of $n_0$ and $n_2$, so the squeezing will help in getting a larger mean-photon number.  Similar to the Section.~\ref{sec:hilbert_space}, the mixer was chosen to be $H_m = \sum_i~(x_i-x_0)^2 + (p_i-p_0)^2$. For simplicity, we considered a four qumode circuit but it is important to note that the same circuit structure was applied to the other four qumodes as well. To construct the ansatz for our experiment, we considered the operator pool $\mathcal{A}$ obtained from $A_{\lambda}^{(l)}$ with $l=2$. Among many, this pool included the $R$ gate and the $BS$ gate, which are native to the hardware. Thus, we selected these two gates as the building blocks for our circuit ansatz. To enhance the expressive power of the circuit, we incorporated two ancillary qumodes in addition to the two qumodes required for the problem. This allowed us to exploit the full SU(4) transformation. Therefore, the circuit ansatz consisted of a series of $R$ gates applied to four qumodes, followed by $BS$ gates applied to nearest-neighbor qumodes. A graphical representation of the ansatz is shown in Fig.~\ref{fig:experiment}(b).

Regarding the parameterization, each $R$ gate in the circuit had its own independent free parameter. The $BS$ gate possesses a transmissivity angle and a phase angle, as outlined in Table \ref{tab:gates}. However, in our specific experiments, we set the phase angle to zero and treated the transmissivity angle as a free parameter. Consequently, our ansatz was characterized by $Q=7$ free parameters. Due to limited access to the physical hardware, we ran the optimization loop as a numerical simulation with a cutoff $D=3$ and implemented the circuit with the optimized parameters on the chip to obtain measurement outcomes from 1000 shots. For the optimization, we implemented a $p=1$ layer circuit with a gradient-free optimizer called COBYLA~\cite{Powell1994}. 

Fig.~\ref{fig:experiment}(c) illustrates the mean photon numbers obtained from both the simulator and the experimental setup for all qumodes. Remarkably, even with a moderate value of $D=3$, we successfully obtained the exact solution to the problem. In the numerical simulation, the mean photon numbers for qumodes $(0,1,2,3)$ would be exactly the same as qumodes $(4,5,6,7)$ since the operations are identical. However, in the experimental setting, the distribution deviates due to inherent limitations such as the finite number of shots, noise, and losses in the chip. Additionally, the utilization of a low cutoff in the numerical simulation may have resulted in suboptimal parameters for the actual chip. Despite these factors, a notable resemblance is observed between the experimental results and the numerical simulations. To further analyze the agreement, we computed the average of the mean photon numbers for identical qumodes  $(0,4)$, $(1,5)$. $(2,6)$, and $(3,7)$ and the results are shown in Fig.~\ref{fig:experiment}(d). Encouragingly, this analysis demonstrates a high level of concurrence between the experimental outcomes and the numerical simulations. Therefore, the PCQO algorithm provides a promising circuit ansatz that can be readily implemented using currently available hardware.

It is important to note that the implemented problem possesses relative simplicity as the chip exclusively incorporates fixed squeezing and lacks displacement operations. When encountering problems with large integer solutions, achieving the desired mean photon number becomes challenging without variable squeezing and displacement. Our current experimental results focus on demonstrating the feasibility of the PCQO algorithm through simple experiments. However, future advances enabling variable squeezing or displacement operations will facilitate tackling more complex problem instances. 
\section{Discussions and future work}\label{sec:discussion}
We proposed a hybrid quantum-classical optimization algorithm for photonic quantum computing to tackle complex problems with the currently available technologies. The circuit ansatz for this algorithm is a problem-inspired ansatz computed by utilizing shortcuts-to-adiabaticity techniques, specifically counterdiabatic protocols. We investigated the performance of the algorithm by considering two non-convex continuous-variable optimization problems up to four variables and with a degree of six. We also considered two integer programming problems, specifically the unbounded knapsack problem for up to four system sizes, and the max-clique problem for up to six node graphs. We observed that the PCQO algorithm successfully finds good approximate solutions to these problems using a few gates. To showcase the practical feasibility of PCQO, we conducted experiments on an eight-mode nanophotonic chip. These experiments substantiated that PCQO can be implemented on near-term photonic chips, thereby providing a promising avenue for utilizing photonic quantum computing to solve optimization problems.

We have considered relevant optimization problems for the industry, but this algorithm can be extended to study physical problems as well. PCQO is a hybrid algorithm but purely quantum counterdiabatic algorithms can be developed in the future to study the performance from a point of view of shortcuts-to-adiabaticity. The backend for this algorithm is a photonic system but this can be extended to any bosonic systems as well and the performance analysis would be interesting in this regard. Advanced machine learning techniques like reinforcement learning~\cite{PhysRevX.11.031070} and adaptive techniques~\cite{Grimsley2019} are other aspects that can be incorporated in PCQO to select the circuit ansatz in a better way. Also, finding effective initialization strategies would be interesting for future work. We believe that this work will serve as a benchmark for designing more advanced hybrid qubit-bosonic optimization algorithms~\cite{9926318}. In summary, this work introduces the PCQO algorithm as a compelling approach for addressing hard optimization problems using photonic quantum computing. The successful application of PCQO to various problem domains, combined with its potential for further advancements and extensions, positions photonic quantum computing as a competitive candidate alongside qubit-based technologies for tackling challenging optimization tasks.
\begin{acknowledgments}
We acknowledge the use of Strawberryfields Library~\cite{strawberryfields} for performing the simulations and the experiment. The authors acknowledge Tasio Gonzalez-Raya, Narendra Hegade, and Martin Larocca for useful discussions. This work is supported by EU FET Open Grant  EPIQUS (899368), and the Basque Government through Grant No. IT1470-22, the project grant PID2021-126273NB-I00 funded by MCIN/AEI/10.13039/501100011033 and by ``ERDF A way of making Europe" and ``ERDF Invest in your Future", the Spanish CDTI through Plan complementario Comunicaci\'on cu\'antica (EXP. 2022/01341)(A/20220551), and project OpenSuperQ+100 (101113946) of the EU Flagship on Quantum Technologies, and the IKUR Strategy under the collaboration agreement between Ikerbasque Foundation and BCAM on behalf of the Department of Education of the Basque Government. M.S. acknowledges support from Spanish Ram\'on y Cajal Grant RYC-2020-030503-I. X.C. acknowledges ayudas para contratos Ram\'on y Cajal–2015-2020 (RYC-2017-22482). MGdA acknowledges support from the UPV/EHU and TECNALIA 2021 PIF contract call.
\end{acknowledgments}

\bibliography{reference.bib}
\end{document}